\documentclass[prc,twocolumn,superscriptaddress,showpacs,amssymb,amsmath,amsfonts,aps]{revtex4} 
\usepackage{graphicx} 
\begin{document} 
\title{{\large Cross Sections for the $\gamma p \to K^{*0}\Sigma^+$ 
Reaction at $E_\gamma =1.7-3.0$ GeV  }} 
\preprint{DRAFT: for CLAS Collaboration Review} 
\newcommand*{\kstar}{$K^{*0}$}

\newcommand*{\OHIOU}{Ohio University, Athens, Ohio  45701} 
\affiliation{\OHIOU} 
\newcommand*{\ASU}{Arizona State University, Tempe, Arizona 85287-1504} 
\affiliation{\ASU} 
\newcommand*{\UCLA}{University of California at Los Angeles, Los Angeles, California  90095-1547} 
\affiliation{\UCLA} 
\newcommand*{\CMU}{Carnegie Mellon University, Pittsburgh, Pennsylvania 15213} 
\affiliation{\CMU} 
\newcommand*{\CSU}{California State University, Dominguez Hills 
, Carson, CA 90747 } 
\affiliation{\CSU} 
\newcommand*{\CUA}{Catholic University of America, Washington, D.C. 20064} 
\affiliation{\CUA} 
\newcommand*{\SACLAY}{CEA-Saclay, Service de Physique Nucl\'eaire, F91191 Gif-sur-Yvette, France} 
\affiliation{\SACLAY} 
\newcommand*{\CNU}{Christopher Newport University, Newport News, Virginia 23606} 
\affiliation{\CNU} 
\newcommand*{\UCONN}{University of Connecticut, Storrs, Connecticut 06269} 
\affiliation{\UCONN} 
\newcommand*{\ECOSSEE}{Edinburgh University, Edinburgh EH9 3JZ, United Kingdom} 
\affiliation{\ECOSSEE} 
\newcommand*{\FIU}{Florida International University, Miami, Florida 33199} 
\affiliation{\FIU} 
\newcommand*{\FSU}{Florida State University, Tallahassee, Florida 32306} 
\affiliation{\FSU} 
\newcommand*{\GWU}{The George Washington University, Washington, DC 20052} 
\affiliation{\GWU} 
\newcommand*{\ECOSSEG}{University of Glasgow, Glasgow G12 8QQ, United Kingdom} 
\affiliation{\ECOSSEG} 
\newcommand*{\ISU}{Idaho State University, Pocatello, Idaho 83209} 
\affiliation{\ISU} 
\newcommand*{\INFNFR}{INFN, Laboratori Nazionali di Frascati, 00044 Frascati, Italy} 
\affiliation{\INFNFR} 
\newcommand*{\INFNGE}{INFN, Sezione di Genova, 16146 Genova, Italy} 
\affiliation{\INFNGE} 
\newcommand*{\ORSAY}{Institut de Physique Nucleaire ORSAY, Orsay, France} 
\affiliation{\ORSAY} 
\newcommand*{\ITEP}{Institute of Theoretical and Experimental Physics, Moscow, 117259, Russia} 
\affiliation{\ITEP} 
\newcommand*{\JMU}{James Madison University, Harrisonburg, Virginia 22807} 
\affiliation{\JMU} 
\newcommand*{\KYUNGPOOK}{Kyungpook National University, Daegu 702-701, South Korea} 
\affiliation{\KYUNGPOOK} 
\newcommand*{\UMASS}{University of Massachusetts, Amherst, Massachusetts  01003} 
\affiliation{\UMASS} 
\newcommand*{\MOSCOW}{Moscow State University, General Nuclear Physics Institute, 119899 Moscow, Russia} 
\affiliation{\MOSCOW} 
\newcommand*{\UNH}{University of New Hampshire, Durham, New Hampshire 03824-3568} 
\affiliation{\UNH} 
\newcommand*{\NSU}{Norfolk State University, Norfolk, Virginia 23504} 
\affiliation{\NSU} 
\newcommand*{\ODU}{Old Dominion University, Norfolk, Virginia 23529} 
\affiliation{\ODU} 
\newcommand*{\RPI}{Rensselaer Polytechnic Institute, Troy, New York 12180-3590} 
\affiliation{\RPI} 
\newcommand*{\RICE}{Rice University, Houston, Texas 77005-1892} 
\affiliation{\RICE} 
\newcommand*{\PITT}{University of Pittsburgh, Pittsburgh, Pennsylvania 15260} 
\affiliation{\PITT} 
\newcommand*{\URICH}{University of Richmond, Richmond, Virginia 23173} 
\affiliation{\URICH} 
\newcommand*{\SCAROLINA}{University of South Carolina, Columbia, South Carolina 29208} 
\affiliation{\SCAROLINA} 
\newcommand*{\JLAB}{Thomas Jefferson National Accelerator Facility, Newport News, Virginia 23606} 
\affiliation{\JLAB} 
\newcommand*{\UNIONC}{Union College, Schenectady, NY 12308} 
\affiliation{\UNIONC} 
\newcommand*{\VT}{Virginia Polytechnic Institute and State University, Blacksburg, Virginia   24061-0435} 
\affiliation{\VT} 
\newcommand*{\VIRGINIA}{University of Virginia, Charlottesville, Virginia 22901} 
\affiliation{\VIRGINIA} 
\newcommand*{\WM}{College of William and Mary, Williamsburg, Virginia 23187-8795} 
\affiliation{\WM} 
\newcommand*{\YEREVAN}{Yerevan Physics Institute, 375036 Yerevan, Armenia} 
\affiliation{\YEREVAN} 
\newcommand*{\NOWRU}{Rochester University, Rochester, New York 14627} 
\newcommand*{\NOWREGINA}{University of Regina, Regina Saskatchewan, Canada S4S 0A2} 
\newcommand*{\NOWUNH}{University of New Hampshire, Durham, New Hampshire 03824-3568} 
\newcommand*{\NOWCMU}{Carnegie Mellon University, Pittsburgh, Pennsylvania 15213} 
\newcommand*{\NOWSACLAY}{CEA-Saclay, Service de Physique Nucl\'eaire, F91191 Gif-sur-Yvette,Cedex, France} 
\newcommand*{\NOWUCLA}{University of California at Los Angeles, Los Angeles, California  90095-1547} 
\newcommand*{\NOWUMASS}{University of Massachusetts, Amherst, Massachusetts  01003} 
\newcommand*{\NOWMIT}{Massachusetts Institute of Technology, Cambridge, Massachusetts  02139-4307} 
\newcommand*{\NOWCUA}{Catholic University of America, Washington, D.C. 20064} 
\newcommand*{\NOWGEISSEN}{Physikalisches Institut der Universitaet Giessen, 35392 Giessen, Germany} 
 
\author {I.~Hleiqawi}  
\affiliation{\OHIOU} 
\author {K.~Hicks}  
\affiliation{\OHIOU} 
\author {D.S.~Carman}  
\affiliation{\OHIOU} 
\affiliation{\JLAB} 
\author {T.~Mibe}  
\affiliation{\OHIOU} 
\author {G.~Niculescu}  
\affiliation{\JMU} 
\author {A.~Tkabladze}  
\affiliation{\GWU} 
\author {M.~Amarian}  
\affiliation{\ODU} 
\author {P.~Ambrozewicz}  
\affiliation{\FIU} 
\author {M.~Anghinolfi}  
\affiliation{\INFNGE} 
\author {G.~Asryan}  
\affiliation{\YEREVAN} 
% ...
\collaboration{The CLAS Collaboration} 
     \noaffiliation 
\date{\today} 
 
\begin{abstract} 
Differential cross sections for the reaction 
$\gamma p \to K^{*0} \Sigma^+$ 
are presented in photon energy range from 1.7 to 3.0 GeV.  
The \kstar~was detected by its decay products, 
$K^+\pi^-$, in the CLAS detector at Jefferson Lab. 
These data are the first \kstar~photoproduction cross sections 
ever published over a broad range of angles.  Comparison with 
a theoretical model based on the vector and tensor $K^*$-quark 
couplings shows good agreement with the data,
except at forward angles, suggesting that the role of scalar 
$\kappa$ meson exchange should be investigated.
\end{abstract} 
 
\pacs{13.60.-r,13.60.Le,13.60.Rj,14.40.Ev} 
 
\maketitle 
 
\section{ Introduction } 
 
The photoproduction of the vector meson $K^*$ has some advantages 
over its non-strange partners, the $\rho$ and $\omega$, 
in probing the dynamics usually included in theoretical calculations. 
Because the $K^*$ has non-zero strangeness, pomeron exchange is 
not possible.  In comparison, $t$-channel pomeron exchange 
dominates the total cross section for $\rho$, $\omega$ and $\phi$
photoproduction at high energy. Near threshold energies, scalar meson 
exchange in the $t$-channel contributes strongly to $\rho$ and 
$\omega$ photoproduction, but the equivalent diagram for $K^*$
photoproduction, via scalar $\kappa$ exchange, has not been established.  
The $\kappa$(800) was omitted
\footnote{See the notes on scalar mesons in the Meson Particle Listings 
for $f_0(600)$ in Ref. \cite{pdg}.}
from the Particle Data Group \cite{pdg} 
primary tables because the existence of this state is controversial.
Instead, $K^*$ photoproduction  models use $t$-channel kaon exchange, 
constrained by $KN$ scattering and kaon photoproduction data.  
Since this is the dominant term at forward angles, the $t$-channel 
parameters in theoretical models are already known.
However, the lack of $K^*$ photoproduction data in the past has 
prevented further theoretical development.  In particular, the 
need for possible $\kappa$ exchange is uncertain.

We present here the first data for $K^{*0}\Sigma^+$ 
photoproduction from the proton over a wide angular range 
for photon energies near threshold ($E_\gamma = 1.7-3.0$ GeV). 
Because of the short lifetime of $K^{*0} \to K^+ \pi^-$,
a detector with large acceptance is needed to capture the 
decay particles.  Here, we use the CLAS detector \cite{mecking} 
where multiple particles can be detected with reasonable efficiency.

In addition to investigating the $t$-channel, as described above, 
our data will be compared with a theoretical model to extract 
information on the vector and tensor coupling strength of the 
\kstar~to quarks in $s$-channel nucleon resonances.  Nucleon 
resonances are predicted to have significant branching ratios to 
kaons and, if the mass is high enough, to $K^*$ channels 
\cite{capstick}.  The overlap between the quark wave functions, 
as predicted in theoretical models, and the branching ratio for 
nucleon resonances decaying to $K^*Y$, can be calculated using an 
effective theory as described below.  
Comparison of $K^*$ production data with these calculations 
provides feedback for models of nucleon resonances.


In the theoretical description of $\gamma N \to K^* Y$ reactions, 
where $Y$ denotes a hyperon, a simplification occurs for neutral 
\kstar~production, where \kstar~exchange in the $t$-channel is 
strongly suppressed \cite{zhao}
(it requires a higher multipolarity than the dominant $M1$ transition).  
Also there is no seagull diagram, since this term is directly 
proportional to the charge of the vector meson. 
Hence, there are some theoretical simplifications when studying 
\kstar~photoproduction compared with $K^{*+}$ production.

Calculations are available based on a quark 
model description of the meson-hyperon vertex in the $s$ and $u$ 
channels \cite{zhao}. In this model there are only two free parameters 
corresponding to the vector and tensor couplings, which are common 
to all intermediate states ($N^*$ or $Y^*$) and depend only on the 
quark mass.  Using this approach, a consistent picture of both 
$K^*$ and non-strange vector meson production can be developed. 
In Ref. \cite{zhao}, the cross sections for $K^*$ production were 
predicted based on SU(3) symmetry and quark coupling parameters 
extracted from $\omega$ photoproduction data.  
The current data will be compared with this theoretical model.

\begin{figure} 
\includegraphics[scale=0.4]{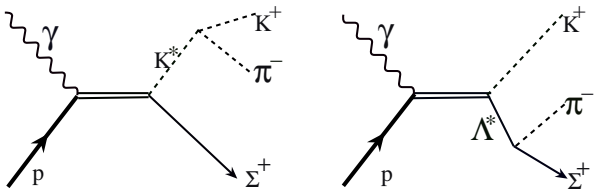} 
\caption{\label{diagrams} 
Schematic diagrams showing \kstar~(left) and $Y^*$ (right) production.
Both diagrams lead to the same final state.
} 
\end{figure}

In contrast to the theoretical advantages of \kstar~photoproduction, 
there are some experimental challenges to be met.
The cross sections are small, in part due to the larger mass of the 
$K^*$, which suppresses production compared to that of lighter mesons.  
Also, the multiparticle final state 
requires large acceptance detectors rather than traditional 
magnetic spectrometers with small solid angle acceptances.
Perhaps for this reason, there is only one experiment 
reported in the literature for $K^*$ photoproduction 
\cite{atkinson}, using 20-70 GeV photons (well above threshold).
More troublesome is the physics background from production of $Y^*$ 
hyperon resonances which decay to $Y\pi$ final states; 
these reactions have overlapping kinematics with $K^*$ production 
and must be subtracted in the analysis. 
An example of two diagrams having the same final state 
is shown in Fig. \ref{diagrams}.  
As discussed below, a Dalitz plot of the invariant mass of the 
$K\pi$ system versus that of the $Y\pi$ system (not shown) has 
horizontal and vertical bands for $K^*$ and $Y^*$ resonances, 
respectively.  The bands will cross where the kinematics overlap, 
at which point these reactions are indistinguishable. 
As described in Section III, 
care must be taken to extract the \kstar~cross sections independently 
from the $Y^*$ background. 

From a broader perspective, data for $(\gamma,K)$ and $(\gamma,K^*)$ 
reactions together provide clear information about the production of 
an $s\bar{s}$ pair from the vacuum \cite{zhao}.  For example, measurements 
of polarization transfer in the $\vec{e} p \to e'K^+\vec{\Lambda}$ 
reaction \cite{carman} 
have shown that the $s\bar{s}$ are produced with spins primarily 
anti-aligned.  In contrast, similar measurements with the spin-1 $K^*$ 
are expected to produce the $s\bar{s}$ pair with 
spins aligned.  A coherent description of both reactions will 
likely be needed in order to do a full coupled-channel analysis, 
since $K^*$ exchange contributes to $t$-channel diagrams in 
kaon production, and kaon exchange contributes to $t$-channel 
diagrams in $K^*$ production \cite{hlee}.

\section{ Experimental Details }

The experiment was carried out in Hall B of the Thomas Jefferson 
National Accelerator Facility, using the CLAS detector \cite{mecking}. 
An electron beam of energy 3.115 GeV was incident on a gold foil, 
creating a bremsstrahlung photon beam.  Each photon 
was tagged by measuring the energy of the electron in a tagging 
spectrometer \cite{sober}, with an upper range at 95\% of the 
incident electron energy. 
The photons then passed through a collimator before reaching 
an 18.0 cm long liquid-hydrogen target in the center of CLAS.  
The target was surrounded by a segmented scintillator, 
called the start counter \cite{taylor}. 

The momentum of charged particles was determined from tracks in the 
drift chambers \cite{mestayer}, located outside of the start counter. 
CLAS has a toroidal magnetic field \cite{mecking}, which was set at 
50\% of the maximum and oriented so that negatively charged particles 
bent inward toward the beam line. 
The outer layer of the drift chambers surrounds the region of 
strongest magnetic field.  
The time-of-flight (TOF) of particles was measured between the stop 
time from scintillators outside the drift chambers \cite{elton} 
and the start counter time.  A combination 
of the momentum (from the drift chambers) and the TOF (from the 
scintillators) was used to calculate the mass of each particle. 
A coincidence between the tagger, the start counter and a TOF 
scintillator provided the event trigger. 

We used the $\Delta \beta$ method for particle identification, 
where the measured velocity (from the TOF) is subtracted from 
the velocity calculated from the momentum for a given mass, 
and values of $|\beta_{TOF} - \beta_{calc} |<0.1$ were accepted.
The particle vertex time is calculated from the TOF for a given mass 
(of the kaon or pion) along with the momentum measured by the 
drift chambers. The particle vertex time is required to be  
within 1.0 ns of the photon time, measured by the tagger and 
extrapolated to the vertex.
In addition, our analysis required both a positive kaon and a negative 
pion to be identified in coincidence, which suppressed accidental 
coincidences considerably.  The same analysis steps were used 
for both data and Monte Carlo simulations, so particle 
misidentification due to kaon decays could be accounted for.

The measurement of the photon flux has been described previously 
\cite{pasyuk}. Briefly, the main idea is to first count the number 
of ``good" electrons in the tagging spectrometer, given by a 
timing coincidence of scintillators at the tagger focal plane, 
and comparing this with the number of counts in a total absorption 
counter (four blocks of lead glass) placed directly in the photon 
beam during low-intensity calibration measurements.  
Recoiling electrons both in and out of a coincidence window were 
monitored and used to calculate the rate of photons produced at 
the tagger.  The tagger was upstream of the target, so this rate 
was corrected for loss of photons in the beamline between the 
tagger and the physics target.  The rate of photons striking the 
target was scaled by the experimental livetime.

Monte Carlo simulations of the detector were carried out using 
the software package GEANT \cite{geant}, input with the detector 
configuration of CLAS.  Detector acceptance factors obtained 
from simulations have been shown previously to reproduce the 
world data on known reactions \cite{mcnabb}. 
Two event generators were used.  One has a uniform distribution 
in two-body phase space for $K^{*0}\Sigma^+$ production, and the 
other has angular distributions calculated from the model of 
Zhao \cite{zhao} for our reaction.  The detector acceptances 
obtained using these event generators were then used to 
calculate the cross sections, and systematic uncertainties in the 
acceptance factor were estimated by comparing the two results.
The average acceptance for \kstar~detection in CLAS was about 3\% 
and ranges from 1-6\% depending on the photon beam energy and 
the \kstar~angle.  Details are given in Ref. \cite{ishaq}.

\section{ Data Analysis }

Events were scanned for $K^+$ and $\pi^-$ candidates, based on the 
particle identification method described in the previous 
section.  Both particles were extrapolated in time to the 
target vertex and were required to be within 1.0 ns of the 
photon time.  The time resolution of the TOF 
system, $\sigma_t$, is about 100-300 ps, depending on the location 
in the CLAS detector. A time cut of $\pm 3\sigma_t$ was used for 
each particle. The time resolution in the Monte Carlo simulation 
is well matched with that seen by our experiment, 
so the loss of particles 
(for example due to $K^+$ decay) outside of the timing cuts is 
folded into the detector acceptance calculation.

\begin{figure}[htb]
\includegraphics[scale=0.45]{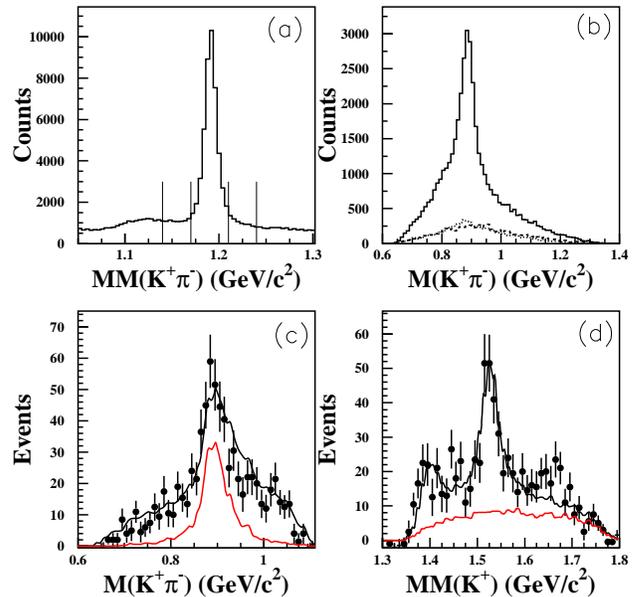} 
\caption{\label{massfits} (Color online) 
(a) Missing mass for the $\gamma p \to K^+\pi^- X$ reaction  
summed over all angles and photon energies. 
(b) Invariant mass of the $K^+\pi^-$ system, with the 
same event selection, for events in the $\Sigma^+$ 
peak (solid) and the sideband regions (dashed and dotted) 
shown in (a).
(c) $M(K^+\pi^-)$ for $E_\gamma=$2.24-2.36 GeV and 
$-0.2<\cos\theta_{K^*}<0.1$ showing the full fit (black)
and the \kstar~peak component (red).
(d) $MM(K^+)$ showing the $\Lambda$(1520) and other hyperon peaks 
(see Fig. 1, right) along with the curves as in (c).
} 
\end{figure}

After identifying a valid $K^+$ and $\pi^-$, 
the missing mass for the $\gamma p \to K^+ \pi^- X$ reaction, 
$MM(K^+\pi^-)$, was calculated, as shown by Fig. \ref{massfits}(a).   
The $\Sigma^+$ peak dominates the 
spectrum, at a mass of 1.189 GeV/c$^2$, riding on a smooth background.
Next, the invariant mass of the system, $M(K^+\pi^-)$ is calculated.  
After requiring events in the $\Sigma^+$ region, 
$1.175 < MM(K^+\pi^-) < 1.205$ GeV/c$^2$, the invariant mass spectrum 
is shown in Fig. \ref{massfits}(b).
The \kstar~peak is visible at a mass of 0.896 GeV/c$^2$, 
which has an intrinsic width of about 0.05 GeV/c$^2$.  
The \kstar~peak sits on top of background due to competing 
processes, such as $Y^*$ production (see Fig. \ref{diagrams}).  
The non-$Y^*$ background can be removed by the sideband 
subtraction method.  Events in regions of $MM(K^+\pi^-)$ on 
either side of the $\Sigma^+$ region, inside the vertical 
lines in Fig. \ref{massfits}(a), are plotted by the 
dashed and dotted lines in Fig. \ref{massfits}(b).

Background from $\Sigma (1385)$, $\Lambda(1405)$ and $\Lambda(1520)$ 
resonances were calculated by simulations with angular distributions 
described by an exponential $t$-slope \cite{barrow},   
which are then projected onto the $M(K^+\pi^-)$ mass.  The 
strength of each $Y^*$ resonance is treated as a free 
parameter in the fits to each kinematic bin (see below).
Interference between $Y^*$ resonances and \kstar~production 
were studied and found to be small \cite{ishaq}, keeping 
the cross sections within the systematic uncertainties.

The data were processed in 6 angular bins for each of 
9 photon energy bins.  The variable for the angular bins 
was $\cos \theta_{K^*}$ (where $\theta_{K^*}$ 
is the angle of the \kstar~in the center of mass system). 
Fewer counts were found at backward angles, and so the 
bin sizes were made larger for $\cos \theta_{K^*} < 0$.
Note that our angular coverage stops at 
$\cos \theta_{K^*} = 0.9$.  Beyond this point, the detector 
acceptance falls off sharply.
The photon energy bins span the range from 1.71 GeV up to 
2.96 GeV. The photon energy bins are not equally spaced 
in order to avoid regions where tagger counters were not 
working properly.

Yields in each bin were extracted by fitting the \kstar~peak, 
which rides on a nonlinear background.  
The background due to sidebands of the $\Sigma^+$ peak 
were subtracted in each bin.  The remaining background was 
fit directly to the shape of Monte Carlo simulations by adjusting 
the relative magnitudes of 3-body phase space ($K^+$,$\pi^-$,$\Sigma^+$)
and $K^+Y^*$ event generators, where $Y^*$ represents the 
$\Sigma(1385)$, $\Lambda(1405)$ and $\Lambda(1520)$ resonances.
A sample result of the template fits is shown in parts (c) and (d) 
of Fig. \ref{massfits} for a single kinematic bin.
The strengths of the $Y^*$ resonances were constrained by 
simultaneous fits to the missing mass spectra of the kaon, $MM(K^+)$, 
for each bin. 
The shape of the \kstar~peak was fixed to the shape from 
the GEANT simulations, which used a Breit-Wigner parameterization 
for the known \kstar~resonance mass and width.
The reduced $\chi^2$ of the fits was typically between 1.0 to 2.5. 
When the fit is restricted to the regions surrounding the 
\kstar~peak and low-lying $Y^*$ resonances, then values of 
$\chi^2$ near unity are obtained.  This suggests that higher mass 
resonances, like the $\Lambda(1660)$, could improve the overall 
simulation of the data.  However, the fits indicate that the 
addition of more $Y^*$ resonances will not change the cross 
sections outside of the systematic uncertainty range given below.

As a measure of the systematic uncertainty of the fitting 
procedure, we varied several event selection parameters and 
refit the \kstar~peak.  For example, we varied the width of the 
mass cut on the $\Sigma^+$ peak, and the corresponding sideband 
regions.  Also, we tried the fits with and without the $\Lambda(1405)$ 
resonance, which has nearly overlapping kinematics with the 
$\Sigma(1385)$ resonance.  After comparing the fits for variations 
in the analysis procedures, we determined that the uncertainty 
in the \kstar~yield is $\pm$18\%.  

%

The cross section in each kinematic bin depends on the detector 
acceptance.  The CLAS detector has been modeled using 
the software package GEANT, as described above, using 
two different event generators.  Because of the large 
bins in angle, which were necessary due to the limited 
statistics in the \kstar~peaks, there was not much 
difference between acceptances calculated using each 
event generator.  Hence, we chose the flat phase-space 
model for the detector acceptance corrections
rather than the one based on Ref. \cite{zhao}.
The branching ratio of $K^{*0} \to K^+\pi^-$ 
has been accounted for.

The systematic uncertainty in the yield extraction, 
as described above was 18\%.  Other 
systematic uncertainties were calculated from varying 
the cuts used for particle identification (8\%), 
varying the allowed fiducial region for the 
particles detected in CLAS (8\%), the dependence 
on the Monte Carlo event generator (4\%) and the 
luminosity normalization (7\%).  Added in quadrature, 
the total systematic uncertainty for our measurement 
is $\pm$23\%. 

\section{ Results and Discussion }

\begin{figure} 
\includegraphics[scale=0.4]{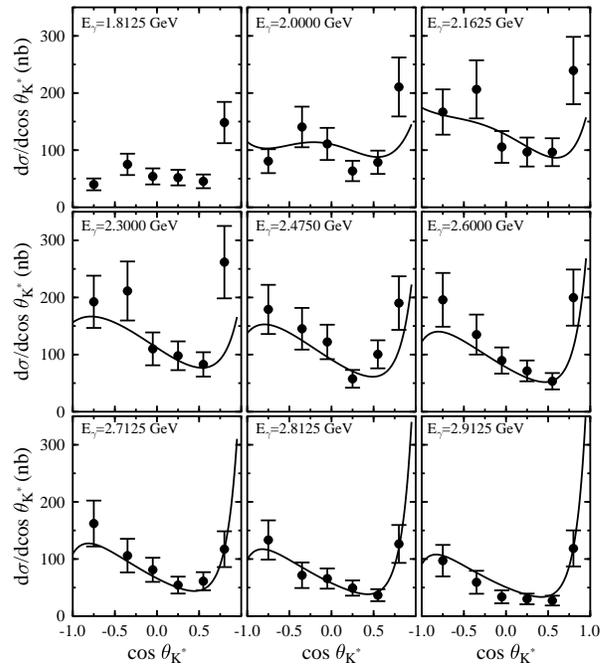} 
\caption{\label{dxdo} Cross sections for the reaction 
$\gamma p \to K^{0*} \Sigma^+$ as a function of angle for 
the energy bins shown.  The curves are from the model 
of Zhao \cite{zhao} using vector and tensor couplings 
adjusted to give the best fit to the data.  Error bars 
include systematic uncertainties.  For numbers, 
see Ref. \cite{arxiv}.
} 
\end{figure}

The differential cross sections are presented in Fig. \ref{dxdo} 
for nine photon energy bins. Also shown 
are curves calculated from the model of Zhao \cite{zhao} using 
the parameters $a=-2.2$ and $b=0.80$ for the vector and tensor 
$K^*$-quark couplings, respectively.  These parameters are 
different than those given in Ref. \cite{zhao}, 
$a=-2.8$ and $b=-5.9$, which were 
extrapolated from $\omega$ production data using SU(3) symmetry. 
The current values, which were fitted to the \kstar~data, are 
significantly smaller than those predicted in Ref. \cite{zhao}, 
however it is worth noting that some corrections were made to the 
computer code used in Ref. \cite{zhao}, so the parameters 
$a$ and $b$ used in that reference should be reexamined. 
The first photon energy bin was not fit because of numerical 
instability of the calculation close to threshold.

Near threshold, the data have a fairly flat angular distribution. 
As the energy increases, the forward and backward angles dominate, 
showing substantial $t$-channel and $u$-channel contributions. 
The $u$-channel is implicitly included in the $K^*$-quark couplings 
(parameters $a$ and $b$ above) which also control the $s$-channel 
amplitudes.  The $t$-channel coupling in the Zhao model is,
in principle, completely determined by a single 
diagram with $K^0$ exchange, where the $KN\Sigma$ vertex coupling 
is taken from $KN$ scattering and the $\gamma K^+ K^*$ electromagnetic 
coupling is fixed from $K^*$ radiative decay.
In other words, there should be no free parameters in the $t$-channel 
for this model.  However, in order to obtain better agreement with 
the data at forward angles, the $t$-channel amplitude was multiplied 
by a factor of 0.8.  This factor is perhaps justified because 
there may be other $t$-channel diagrams that can contribute, 
such as scalar $\kappa$ exchange, 
that are not included in the Zhao model.

\begin{figure} 
\includegraphics[scale=0.4]{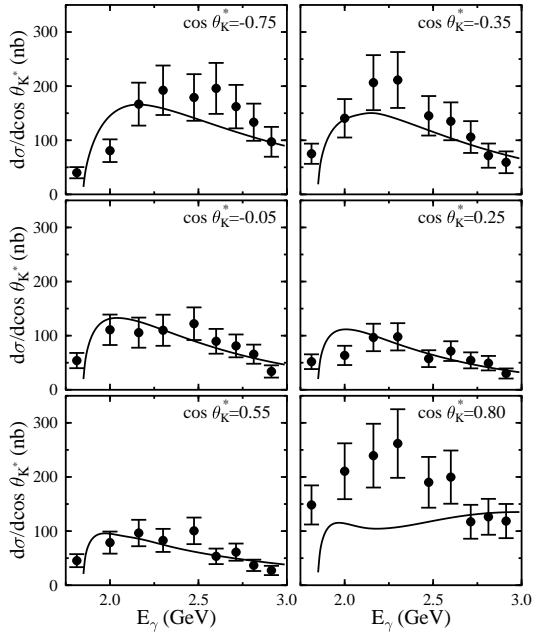} 
\caption{\label{dxde} Cross sections for the reaction 
$\gamma p \to K^{0*} \Sigma^+$ as a function of energy for 
the angular bins shown.  The curves are from the same model 
as in the previous figure.  Error bars include systematic 
uncertainties.
} 
\end{figure}

The energy dependence of the cross sections is presented in 
Fig. \ref{dxde} for the six angle bins shown.  The solid 
curves use the same parameters ($a$ and $b$) as given above. 
The good agreement between theory and experiment suggests 
that the $K^*$-quark dynamics used in the Zhao model are 
reasonable. The only serious discrepancy is for the 
most forward angle bin, where the global fit underpredicts 
the cross section at lower photon energies.  
Clearly, the energy dependence of the $t$-channel strength 
within the context of the Zhao model is not correct to reproduce the 
forward angle strength.  Note that if the Zhao model is fitted 
to this single angular bin, a reasonable fit can be obtained, 
but at the expense of destroying the good agreement with the 
data at larger angles.

\section{ Conclusions }

Using the CLAS detector, the first near-threshold measurement of 
\kstar~photoproduction cross sections has been obtained. 
The results reported here show that the angular distributions 
are forward peaked, as expected from $t$-channel diagrams 
with kaon exchange, and increase slowly at backward 
angles, showing the effect of $s$- and $u$-channel resonances that 
couple to $K^{*0}\Sigma^+$.  As a function of energy, 
the cross sections exhibit a maximum just 
above threshold and then a smooth decrease.

The data are in good agreement with the Zhao model, which 
uses a $K^*$-quark coupling with only two constant parameters,
except at the most forward angle bin.
The forward angles are dominated by the $t$-channel, where
$K^0$ exchange is expected to have the largest amplitude.
By tuning the model parameters, in a global fit to the data, 
the calculations reproduce the general trend of the data 
at larger angles.
The other aspects of this model are constrained by previous 
data or predicted from resonance properties from quark models.
The lack of agreement for the energy dependence of the $t$-channel 
could indicate that other diagrams, like scalar $\kappa$
exchange, need to be investigated. 

Improvements in the statistics will enable smaller kinematic 
bins, which in turn will further constrain comparison with 
theoretical models.  In addition, photoproduction of 
$K^{*+}\Lambda$ will provide new information, and simultaneous 
comparisons of theoretical models to both $K^{*+}$ and 
\kstar~reactions, which have different $t$-channel dependences, 
will shed light on the role of scalar $\kappa$ exchange \cite{oh}. 
Due to parity and angular momentum conservation, the $\kappa$ cannot 
contribute to conventional kaon photoproduction, so data on $K^*Y$ 
reactions is perhaps the best way to investigate the possibility of 
$\kappa$ exchange.  Preliminary data on $K^{*+}$ are already available 
\cite{guo} and larger data sets from CLAS are 
being analyzed \cite{g10,g11}.  


Special thanks go to Q. Zhao and C. Bennholds for discussions and 
access to their computer code.  Event generators for the $Y^*$ states 
are from L.~Guo and G.~Mutchler.
We acknowledge the outstanding efforts of the staff  
of the Accelerator and Physics 
Divisions at Jefferson Lab who made this experiment possible. 
This work was supported in part by the U.S.~Department of Energy, 
the National Science Foundation, 
the Istituto Nazionale di Fisica Nucleare, the French 
Centre National de la Recherche Scientifique and Commissariat \`a 
l'Energie Atomique, and the Korea Science and Engineering Foundation.
Jefferson Science Associates (JSA) operates the 
Thomas Jefferson National Accelerator Facility for the 
U.S.~Department of Energy under contract DE-AC05-060R23177.  


\begin{table}
\caption{ Cross sections, with total uncertainties, in each bin for 
the $\gamma p \to K^{*0} \Sigma^+$ reaction. }
\begin{tabular}{|cc||ccccccc|}
\hline
 \multicolumn{2}{|c||}{ Energy bin }& & \multicolumn{6}{c|}{ Cross section (nb) for each kinematic bin } \\
 \multicolumn{2}{|c||}{ (GeV) } & cos$\theta_{K^*}$ & $-0.75$ & $-0.35$ & $-0.05$ & 0.25 & 0.55 & 0.8 \\
 $E_\gamma$ & $\Delta E_\gamma$ & $\Delta$cos$\theta_{K^*}$ & 0.5 & 0.3 & 0.3 & 0.3 & 0.3 & 0.2 \\
\hline
\hline
 1.8125& 0.2000 & & 40.1$\pm$10.3 & 75.1$\pm$18.7 & 54.0$\pm$14.1 & 51.9$\pm$13.6 & 45.4$\pm$12.0 & 148.3$\pm$36.1 \\
 2.0000& 0.1750 & & 80.7$\pm$21.0 & 140.6$\pm$35.5 & 110.9$\pm$28.1 & 63.5$\pm$17.8 & 78.7$\pm$20.4 & 210.6$\pm$51.6 \\
 2.1625& 0.1500 & & 166.7$\pm$39.7 & 206.4$\pm$50.8 & 105.6$\pm$27.8 & 96.7$\pm$25.4 & 96.4$\pm$24.4 & 239.4$\pm$58.9 \\
 2.3000& 0.1250 & & 192.3$\pm$45.7 & 211.4$\pm$51.7 & 110.0$\pm$28.8 & 97.9$\pm$25.2 & 82.8$\pm$21.3 & 261.8$\pm$63.4 \\
 2.4750& 0.1250 & & 179.0$\pm$43.1 & 145.2$\pm$36.5 & 122.1$\pm$30.2 & 57.5$\pm$15.6 & 100.5$\pm$24.6 & 190.1$\pm$46.9 \\
 2.6000& 0.1350 & & 195.8$\pm$47.1 & 135.0$\pm$35.0 & 89.6$\pm$22.9 & 71.5$\pm$18.2 & 53.3$\pm$14.3 & 199.7$\pm$49.2 \\
 2.7125& 0.1000 & & 162.1$\pm$40.2 & 105.9$\pm$29.4 & 81.1$\pm$21.1 & 54.4$\pm$14.8 & 61.1$\pm$15.9 & 117.2$\pm$31.3 \\
 2.8125& 0.1000 & & 133.3$\pm$34.4 & 71.6$\pm$22.4 & 65.8$\pm$17.7 & 49.2$\pm$13.4 & 36.7$\pm$10.5 & 126.4$\pm$33.1 \\
 2.9125& 0.1000 & & 97.1$\pm$27.6 & 59.3$\pm$20.1 & 33.9$\pm$11.3 & 30.1$\pm$9.3 & 27.2$\pm$8.6 & 118.5$\pm$31.6 \\
\hline
\end{tabular}
\end{table}

\end{document}